\documentclass[twocolumn,amsmath,amssymb]{revtex4}
\usepackage{amssymb}
\usepackage{amsmath}
\usepackage{amscd}
 \begin{document}
                                                                                
\title{Macroscopic Multi-Species Entanglement near Quantum Phase Transitions}
\author{ V. Subrahmanyam}
\address{ Department of Physics, Indian Institute of Technology, Kanpur-208016, India}
\date{\today} 
\begin{abstract}
Multi-Species entanglement,  defined for a many-particle system as the entanglement between 
different species of particles,  is shown to exist in the thermodynamic limit of the system size going to infinity. This macroscopic entanglement, as it can exhibit singular behavior, is capable of tracking quantum phase transitions. The entanglement  between up and down spins has been analytically calculated  for the one-dimensional Ising model in a transverse magnetic field.  As the coupling strength is varied, the first derivative of the entanglement shows a jump discontinuity and the second derivative diverges near the quantum critical point.
\end{abstract}
\maketitle
 \noindent


Entanglement in a quantum state is a signature of quantum correlations between different
parts of the system.  Quantum entanglement,  perceived as a resource for quantum communication
and information processing,  has emerged over the last few years
as a major research area in
various diverse fields such as physics, mathematics, chemistry, electrical
engineering and computer science 
\cite{Nielsen, Benenti}. 
The von Neumann entropy of
a subsystem is a measure of the entanglement between the subsystem and the rest. Mostly, the
entanglement between two spatially separate and distinct parts has been studied in various systems.
In this article we will address the entanglement between two species of particles sharing the same
physical space. As we shall see below, that the thermodynamic limit exists for the two-species entanglement,  i.e. the entropy per volume (size) tends to a finite constant, in the thermodynamic limit of the system size becoming macroscopically large. Thus, the
macroscopic two-species entanglement is a natural candidate for tracking quantum phase transitions.
which will be explored in the context of the transverse-field Ising model of the spin systems, a prototypical system of quantum critical phenomena.

 Let us consider two species of hard-core particles $A$ and $B$ moving on a lattice of $N$ sites, the corresponding
numbers of particles for the species being $N_A$ and $N_B$ respectively. Local hilbert space for
a site is four-dimensional, either $A$ or $B$ occupation, double occupation, and no occupation.  The basis states of the bipartite system can be chosen to be the direct products of the basis states of the individual parts: $|u\rangle_A |v\rangle_B$, where $u$ ($v$) is a set of site locations occupied by  $A$-type ($B$-type) particles. A general pure state of the combined $AB$-system is given by
\begin{equation}
|\psi\rangle = \sum_{u,v} \psi(u,v)~ |u\rangle_A~ |v\rangle_B,
\end{equation}
and the wave function amplitudes $\psi(u,v)$  determine all the properties of the system.
The von Neumann entropy of the subsystem (of particles $A$), the measure of the entanglement
between the two species is  calculated from the eigenvalues  of the
reduced density matrix $\rho_A= Tr_B |\psi\rangle\langle\psi|$. In general the computation of $\rho_A$ is
quite involved,  but it is easier for the case of exclusion and half filling. That is,
the exclusion forbids double occupancy of the two species at a given site,
and the condition of half filling (the total number of particles equals the number of sites) implies that every site is occupied by either $A$ or $B$ particle.
In this case,  the exclusion and half filling conditions map all the physical states of this system to a spin system, each site occupied by a spin-1/2.  Since, each site is either occupied by one of the species, knowing the locations of $A$ particles is enough to characterize the wave functions shown above. Here, $v$ would be just
the complement of $u$. Now, the state can be written as,
\begin{equation}
|\psi\rangle = \sum_{u} \psi(u)~ |u\rangle_A~ |u\rangle_B,
\label{Schmidt}
\end{equation}
where for the state of $B$, the set $u$ denotes  site locations not occupied by $B$ particles.  Now, the above bipartite state already has the desired Schmidt decomposition built in.
The quantum entanglement between the two species is given by the von Neumann entropy,
\begin{equation}
S_A=-\sum_{\vec u} | \psi(u)|^2 \log |\psi(u)|^2.
\end{equation}
This is the desired simplification for the exclusion at half filling, with the wave functions
themselves being the Schmidt numbers.  The entanglement between the two species increases with
the correlation present in the state. In the case of the two species being up and down spin particles of many-electron systems,   a strongly-correlated state (with severe local constraint of no double occupancy) will have larger entanglement  and
uncorrelated spin states (for example metallic states) do not have any entanglement. In contrast,
the usual entanglement measures studied  for  distinct spatial blocks  is maximized if the local constraints are relaxed 
 \cite{Subrahmanyam}.

The above von Neumann entropy is invariant if the basis chosen for $A$ or $B$ particles
is the momentum basis (a unitary transform of the position basis used here), but it  is
not invariant under a SU(2) unitary that can lead to hybrid particles. 
This should be contrasted with the usual entanglement in spin systems explored between two distinct
spatial parts, which is invariant under  a SU(2)  at any site, but not invariant
if the basis is changed to the momentum basis (as this will mix the two partitions). 
 Since both species of particles access the full physical space of states, independent of the
basis, the entanglement between the two species is expected to be extensive with the system size.  If
the density of  particles for both species is finite, 
 there is a possibility of a macroscopic entanglement,in the thermodynamic limit
$N\rightarrow \infty$, defined as
\begin{equation}
\varepsilon_{A,B}= {\rm lim}_{N\rightarrow \infty} {S_A\over N}.
\end{equation}
We will show below that a nonzero entanglement exists in the thermodynamic limit, which will give us a handle for examining the entanglement of quantum phase transitions. In contrast, the  bipartite entanglement between two spacial blocks does not have the thermodynamic limit, as the block entropy does not scale with the system size 
\cite{Vidal, Verstraete}.
A quantum phase transition is usually accompanied by a singular behavior of a thermodynamic observable as the coupling strength is tuned to a critical value, similar to the thermal phase transition where a thermodynamic potential exhibits a singularity as the temperature is tuned to a critical value. We expect the multi-species  entanglement described above is capable of highlighting the quantum transition in the thermodynamic limit. Below, we will study the macroscopic entanglement in the context of the Ising model in the presence of a transverse magnetic field, which is a prototype exactly-solved model of quantum phase transitions. 

Let us consider $N$ lattice sites occupied by a spin-1/2 species, whose interactions are described by the Hamiltonian given by,\begin{equation}
{\cal H}=-J\sum_i \sigma_i^x \sigma_{i+1}^x  - h \sum_i \sigma_i^z.
\end{equation}
Here, $\sigma_i^x, \sigma_i^z$ are pauli operators acting on the local hilbert space at site $i$. Let
$|\uparrow\rangle, |\downarrow\rangle$ be the $\sigma_i^z$ eigenstates with  $\sigma_i^z=\pm 1$
respectively. The system exhibits long-ranged order, either ferromagnetic (for $J>0$) or antiferromagnetic (for $J<0$)
when $h=0$. The ground state is a direct product of $\sigma_i^x$ eigenstates, $|\pm\rangle=|\uparrow\rangle \pm|\downarrow\rangle/\sqrt{2}$, the ferromagnetic state having all sites in $|+\rangle$ state where as the antiferromagnetic state has alternating sites in $|+\rangle$ and $|-\rangle$ states. As the
transverse field is switched on, quantum fluctuations induce excitations, and the ground state is a coherent superposition of $h=0$ ground and 
excited states, with the weights depending on the coupling strength. The situation is similar to inducing transitions to excited states, through coupling to a thermal bath, except that the state of the
system in that case would be an incoherent mixture of states, with the temperature determining the
weights.  For $|J|> h$, the
system exhibits long-ranged order, and for $|J|< h$ there is no long-ranged order. The system
exhibits a quantum critical behavior at $|J|=h$, with a vanishing excitation energy gap
\cite{Sachdev} .
The concurrence measure of entanglement between nearest neighbor sites  is shown to be singular as a function of the
coupling strength, the first derivative showing a logarthmic  divergence
\cite{Osterloh}.
 We will show below that near the quantum
critical points $x\equiv J/h=\pm 1$, the entanglement $\varepsilon(x)$ between up and down spins exhibits singular behavior,  with its derivative showing a jump discontinuity and the its second derivative diverges as $||x|-1|^{-1}$.   
\begin{figure}[t]
\input{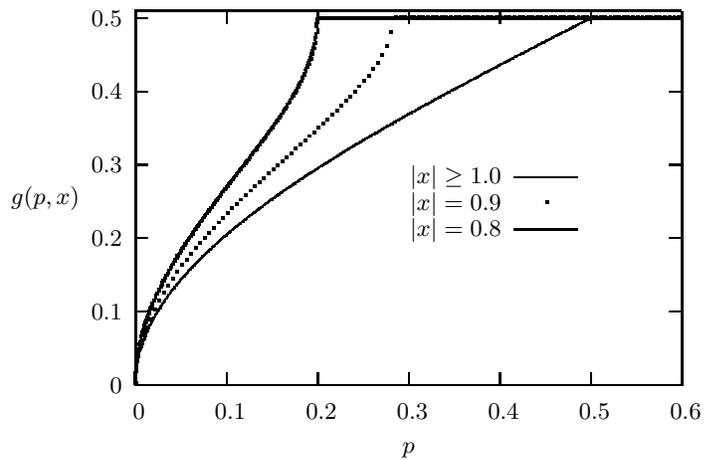}
\caption{The integrated density of eigenvalues as a function of the eigenvalue location, for a few
values of the interaction strength $x=J/h$.}
\end{figure}

\begin{figure*}[t]
 {\hglue -0.8 cm \input{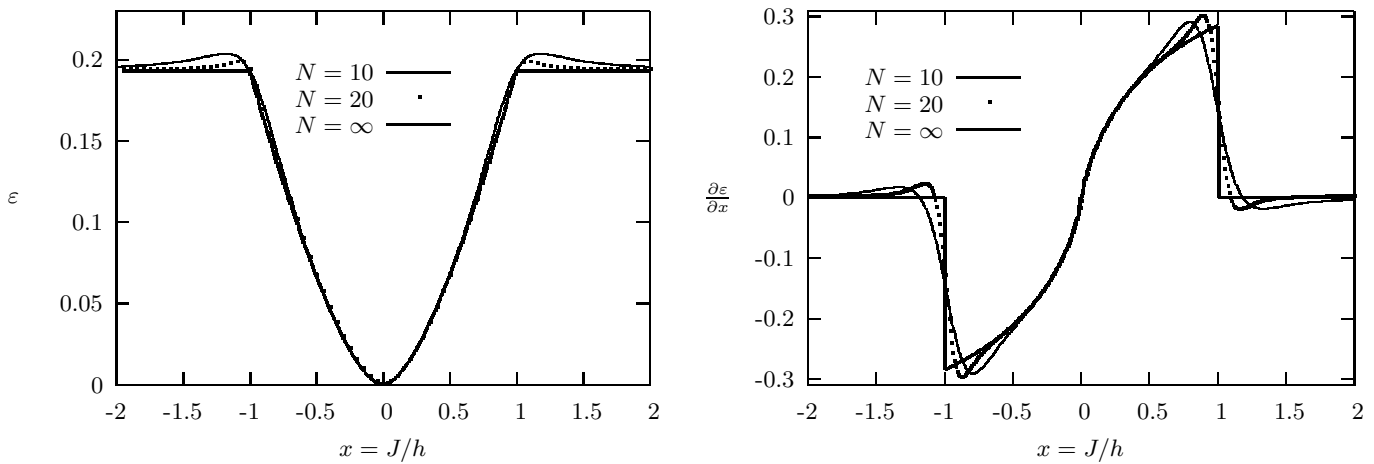}}
\caption{The entanglement, and its   first derivative,  between the up and down spin particles is plotted as a function of the coupling strength, for $N=10,20$,  and in the thermodynamic limit $N\rightarrow \infty$.}
\end{figure*}The above Hamiltonian has been exactly solved for all eigenstates by mapping the spins to fermions
through the Jordan-Wigner transformation
\cite{L-S-M, Pfeuty}.
followed by a Fourier transformation to momentum basis, followed by a Bogoliubov or quadratic-form diagonalization.   The first step is to go from the spin variables  $\sigma_i^z, \sigma_i^x=\sigma_i^++\sigma_i^-$ to  fermion variables $n_i, c_i^\dagger,c_i$ the fermion number operator, the creation and
annihilation operators,  through the Jordan-Wigner transform given as,
\begin{equation}
\sigma_l^z=2n_l-1,~~ \sigma_l^+={\rm e}^{i\pi\sum_{j=1}^{l-1}n_j} c_l^\dagger.
\end{equation}
 The presence of the phase-factor operators in the above makes the multi-spin correlation functions very difficult to calculate, however for computing the
macroscopic entanglement they do not pose a great difficulty. In terms of the fermion
states, the presence of a fermion at a site (i.e. $n_i=1$) implies that site is occupied by an up spin, and
an absence of a fermion implies the site is occupied by a down spin. Now, let us define the momentum-basis fermion operator,  given by
\begin{equation}
c_q={1\over \sqrt{N}}\sum c_l ~{\rm e}^{-iq l} .  
\end{equation}
The allowed values of $q$ for the case of periodic boundary conditions and $N$ even, are
$
q=\pm {\pi \over N}, \pm{3\pi\over N}...\pm {(N-1)\pi\over N},
$
for the sectors with even  number of fermions $N_F$.  Similarly, for $N_F$ odd, the allowed values
are $
q=0,\pm {2\pi\over N}, \pm{4\pi\over N}...\pm{(N-2)\pi\over N}, \pi.
$
The 
Hamiltonian becomes uncoupled in terms of different $|q|$ values, and for each $q>0$, a simple diagonalization from the basis $|0\rangle, |\phi_q\rangle \equiv c_q^\dagger c_{-q}^\dagger|0\rangle,|-q\rangle \equiv c_{-q}^\dagger |0\rangle, |q\rangle \equiv c_q^\dagger |0\rangle$ yields the eigenvalues and eigenstates of the Hamiltonian. Of particular interest is
the ground state, is given as a direct product
\cite{Pfeuty},
\begin{equation}
|G\rangle = \prod_{q>0} \left ( a_q |0\rangle + b_q |\phi_q\rangle \right ).
\end{equation}
The amplitudes depend on the coupling strength, and thus determine the entanglement
in the state. We have,
\begin{equation}
|a_q|^2= {1\over 2} (1- {h +J \cos q \over \sqrt{h^2+J^2+2Jh \cos q}}  ),
\end{equation}
and the other amplitude is given by $|b_q|^2=1-|a_q|^2$.

The above state is a superposition of many-particle momentum basis states, the
amplitude for a given momentum basis state would be a product of wave function amplitudes, either
$a_q$ or $b_q$ appearing for each value of $q$,  corresponding to either the $q$ state is occupied
by fermions (up spins) or the unoccupied by fermions (or equivalently occupied by holes, i.e. down spins). 
Thus each component of the above state can be labeled by $u$ (as in Eq. 2), a set of $q$ values of
the occupied states of the fermion (up-spin particles), owing to the exclusion property, the complement
of the set $u$ would be the occupied $q$ values for the unoccupied (down-spin particles).
Thus, after tracing over the down-spin degrees of freedom,
the up-spin state is obtained as,
\begin{equation}
\rho_\uparrow= \prod a_q^2 |0\rangle \langle 0|  + \sum_q b_q^2 \prod_{q^\prime \ne q}a_{q^\prime}^2 |0,0..\phi_q..0\rangle \langle 0,0..\phi_q..0|+..\end{equation}
Here, the first component is the contribution of the vacuum state, and the second series term is the
contribution from the two-particle sector and so on. The structure of the reduced density matrix is similar even for the maximum energy state (each of the amplitudes $a_q$ is replaced by the amplitude $b_q$ and vice versa),  and hence the same quantum critical behavior  from the view point  of the two-species macroscopic entanglement.  

The von Neumann entropy can be shown to be a sum of independent contributions of $q$ modes,  the individual mode contribution being equal to the Shannon binary entropy
$H(p_i)=-p_i \log p_i - (1-p_i) \log (1-p_i)$, where the eigenvalue is given by $p_i=|a_{q_i}|^2$.
As the coupling strength $x=J/h$ is varied, the band of eigenvalues $p_i$ and the band of eigenvalues 1-$p_i$ spread in width.  For $|x| <1$, as $q_i$ is varied, $!a_q|^2$ is bounded well below the value of 1/2, and similarly  $|b_q|^2$  well above 1/2, with a gap $\Delta(x)$ between the
two bands of eigenvalues. The gap decreases as $x$ is
increased, and finally the gap vanishing at $|x|=1$. For $|x|>1$, the values of $|a_q|^2$ (thus $|b_q|^2$) span the full
range.  A  gap in the individual mode eigenvalue would imply a gap between the largest and the second largest eigenvalues of the reduced density matrix $\rho_\uparrow$ shown above. From the functional relation of the wave functions given in Eq. 10 we can work out the gap,
and we have,
\begin{equation}
\Delta(x)=\sqrt{1-x^2}~\theta(1-|x|),
\end{equation}
which gives a gap exponent of 1/2, i.e. $\Delta\sim (1-|x|)^{1/2}$ as $|x|\rightarrow 1$. 
In contrast, the
particle-hole excitation gap from the ground state to the excited state vanishes linearly
\cite{Sachdev}.

The macroscopic entanglement (from Eq.5) between  up and down spin particles for a given coupling strength $x=J/h$,  is given by
\begin{equation}
\varepsilon (x)= {1\over N}\sum_{i=1}^{N/2} H(p_i )
 = \int_0^1 dp ~g(p,x)\log {1-p\over p}.
\end{equation}
We have converted the sum  into an integral by introducing the integrated density of eigenvalues, i.e.
the number of eigenvalues $p_i$ below a given value of $p$, defined by
\begin{equation}
g(p,x)={1\over N}\sum_i \theta (p-p_i(x)). 
\end{equation}
Here, $p_i(x)=(1-\zeta_i)/2$ is the eigenvalue location,
$\zeta_i=|1+x\cos q|/\sqrt{1+x^2+2x\cos q}$.
Thus, we have transfered the dependence on the coupling strength and the system size to the density of eigenvalues. The integrated density of eigenvalues can be calculated in the thermodynamic limit,
the details will be published elsewhere.  Using $p=(1-\zeta)/2$, we have $ g(p,x)={1\over2}+ \theta(\zeta)\theta(\zeta^2+x^2-1)(\Phi_--\Phi_+)/2\pi$, where
\begin{equation}
\cos \Phi_\pm= {-1+\zeta^2\pm|\zeta|{\rm Sgn}(x)\sqrt{\zeta^2+x^2-1})\over x}~.
\end{equation}
 The integrated density of eigenvalues is plotted in Fig. 1 for a few values of the coupling strength $x$.
For $p>1/2$, it saturates to the value of 1/2 for all values of $x$,  and 
for $|x|\ge 1$, the function $g(p,x)$ has no dependence on $x$!
The entanglement now is given by,
\begin{equation}
\varepsilon(x)={1\over 4 \pi} \int_0^1 d\zeta \log {1+\zeta\over 1-\zeta} ~(\Phi_- -\Phi_+) \theta(x^2+\zeta^2-1).
\end{equation}

We have plotted in Fig. 2 the entanglement and its first derivative as a function of $x=J/h$ from the sum of Eq.12 for $N=10, 20$, along with that of the infinite chain from equation 15. 
The entanglement reaches its maximum value of about $\varepsilon\approx 0.19$ near $|x|=1$.
For the finite-size cases, the entanglement increases a little after $|x|=1$, and quickly falls to this value. Contrast this, with the entanglement of $\log 2$ for the
 ground state with $h=0$,  implying that the limit of $h\rightarrow 0$ is not the same as $h=0$.
 Even for a finite size cluster of $N=10$ sites, one can see in the figure a clear signature of the jump discontinuity
 near the quantum critical points.
 The value of the macroscopic entanglement as $x\rightarrow\infty$ can be  calculated,  as
$\Phi_\pm\rightarrow \pm\zeta$, we get
$
\varepsilon(x\rightarrow\infty)
\approx 0.19.$

The first derivative shows a discontinuity at
$|x|=1$, which corresponding to the quantum critical point, and the second derivative shows a divergence. 
 Since the behavior of
the entanglement depends on the functions $\Phi_\pm$,  and $d\varepsilon/dx$ depends on 
$\Phi_\pm^\prime$,  we can expand these functions near the quantum critical points.
We have the behavior near the critical point $x=!$,  $\Phi_+^\prime\approx -\sqrt{1-\zeta^2}/\zeta$ which does not exhibit any discontinuity, and $\Phi_-^\prime$ exhibits a discontinuity at $x=1$, we
have 
\begin{equation}
{\partial \Phi_-\over \partial x~~}=-{1-x\over |1-x|}{\sqrt{1-\zeta^2}\over \zeta}.
\end{equation}
Thus, for $x\rightarrow 1^+,~\varepsilon^\prime=0$,  but the derivative is nonzero
as $x\rightarrow 1^-$.
The discontinuity $\delta \varepsilon^\prime$ of the derivative of the entanglement at the quantum critical point is given by
\begin{equation}
 \delta\varepsilon^\prime (x=1)
 \approx 0.28.
\end{equation}

Now, the divergence of $d^2 \varepsilon/dx^2$ depends on 
$\partial^2\Phi^{_\pm}/ \partial x^2$.
Near the quantum critical point, 
 $\Phi_+$ has a finite second derivative,   but  $\Phi_-$ shows a singularity,  as shown by,
\begin{equation}
{\partial^2 \Phi_- \over \partial x^2~} \approx {1\over |x-1|} {\sqrt{1-\zeta^2}\over \zeta}.
\end{equation}
Hence,  the second derivative of the entanglement diverges near $x=1$,  as  $\varepsilon^{\prime\prime}\sim |x-1|^{-1}$.

In conclusion, we have shown that the multi-species macroscopic entanglement,  nonzero in the thermodynamic limit,  exhibits singular behaviour near quantum phase transitions. It will be interesting to see if the dynamics starting from a given initial state can also capture signature of critical behaviour, and whether  finite-temperature behavior can be  explored.


\begin{thebibliography}{}

\bibitem{Nielsen} M. A. Nielsen and I. L. Chuang, Quantum Computation
and Quantum Information (Cambridge University Press,
Cambridge, 2000).
\bibitem{Benenti}G. Benenti, G. Casati, and G> Strini, Principles of Quan-
tum Computation and Information (World Scientic, Sin-
gapore, 2007).
\bibitem{Subrahmanyam} V. Subrahmanyam, Lect. Notes Phys. {\bf 802}, 201 (2010); V. Subrahmanyam, Phys. Lett. {\bf A374}, 3151 (2010).
\bibitem{Vidal} G. Vidal, Phys. Rev. Lett {\bf 91}, 147902 (2003). 
\bibitem{Verstraete} F. Verstraete,  M. M.  Wolf, D. Perez-Garcia, J. I. Cirac,  Phys. Rev. Lett. {\bf 96}, 220601 (2006).
\bibitem{Sachdev} S. Sachdev, Quantum Phase Transitions (Cambridge University Press, Cambridge,  1999).
\bibitem{Osterloh} A. Osterloh,  L.  Amico, G. Falci, R.  Fazio,   Nature,  {\bf 416}, 608-610 (2002).
 \bibitem{L-S-M}  E. Lieb, T. Schultz, and D. Mattis, Ann. Phys.  {\bf 16}, 407 (1961)
\bibitem{Pfeuty}P. Pfeuty, Ann. Phys. {\bf 57}, 79 (1970).
 \end{thebibliography}
\end{document}